# The structure of amorphous two-dimensional materials: Elemental monolayer amorphous carbon versus binary monolayer amorphous boron nitride

Yu-Tian Zhang [a§], Yun-Peng Wang [b§], Xianli Zhang [a],
Yu-Yang Zhang [a,c*], Shixuan Du [a,d], Sokrates T. Pantelides [e,a]

a University of Chinese Academy of Sciences and Institute of Physics, Chinese Academy of Sciences, Beijing 100049, China

b Hunan Key Laboratory for Super Microstructure and Ultrafast Process, School of Physics and Electronics, Central South University, Changsha 410083, China

c CAS Center for Excellence in Topological Quantum Computation, University of Chinese Academy of Sciences, Beijing 100049, China

d Songshan Lake Materials Laboratory, Dongguan, Guangdong 523808, China

e Department of Physics and Astronomy and Department of Electrical and Computer Engineering, Vanderbilt University, Nashville, Tennessee 37235, USA

* Email: zhangyuyang@ucas.ac.cn

§ Yu-Tian Zhang and Yun-Peng Wang contributed equally.

ABSTRACT

The structure of amorphous materials has been debated since the 1930's as a binary question: amorphous materials are either Zachariasen continuous random networks (Z-CRNs) or Z-CRNs containing crystallites. It was recently demonstrated, however, that amorphous diamond can be synthesized in either form. Here we address the question of the structure of single-atom-thick amorphous monolayers. We reanalyze the results of prior simulations for amorphous graphene and report kinetic Monte Carlo simulations based on alternative algorithms. We find that crystallite-containing Z-CRN is the favored structure of elemental amorphous graphene, as recently fabricated, whereas the most likely structure of binary monolayer amorphous BN is altogether different than either of the two long-debated options: it is a compositionally disordered "pseudo-CRN" comprising a mix of B-N and noncanonical B-B and N-N bonds and containing "pseudocrystallites", namely honeycomb regions made of noncanonical hexagons. Implications for other non-elemental 2D and bulk amorphous materials are discussed.

Amorphous materials, unlike their crystalline counterparts, have no long-range order. They have rich applications: e.g., hydrogenated amorphous silicon in photovoltaics and thin-film transistors in liquid-crystal displays [1]; amorphous $SiO_2$ in microelectronics and chromatography [2,3]; and amorphous metal-oxide semiconductors like indium-gallium-zinc oxide (IGZO) in thin-film transistors in organic light-emitting diodes [4-6]. Though crystalline materials are accurately characterized by crystallographic techniques, the atomic structure of amorphous materials has been intensely debated.

The main controversy has been whether amorphous solids, particularly glasses, are continuous random networks as proposed by Zachariasen in 1932 [7] (Z-CRN) or CRNs containing crystallites as advocated earlier by Lebedev [8]. The debate became a major scientific East-West contention during the Cold War [9]. By the late 1980's, amorphous materials were widely accepted to be Z-CRNs, but the debate was later revived by new imaging techniques and simulations [10-18]. In 2021, however, large-scale, atomistic, machine-learning-based simulations demonstrated that changing conditions can produce amorphous silicon in either structure [19]. Even more recent papers reported the realization of amorphous diamond either as a crystallite-containing Z-CRN or a Z-CRN [20,21]. Thus, the long-standing controversy appears to be resolved: the structure of bulk amorphous materials can be either of the two distinct options depending on the method of fabrication [22].

Amorphous 2D materials can be synthesized in a variety of ways [23], but the nature of their atomic structure has not received attention. Such investigations have been exclusively on monolayer amorphous carbon (MAC), also called amorphous graphene, and have their own history. In the early 2010's, amorphization of crystalline graphene was achieved by an electron microscope's electron beam [24,25] and by simulations that start with the crystalline form and gradually introduce disorder [26,27], but both these procedures can in principle be stopped to retain crystallites in a CRN or be continued until they produce a fully CRN structure. Alternative simulations that mimic a synthesis process, however, found that different conditions can in principle produce either form of MAC [27,28]. Reanalysis of the resulting structures that were deemed to be Z-CRNs, however, by simply coloring the hexagons, reveals that these structures in fact contain honeycomb nanocrystallites. More recently, Joo *et al.* [29] reported the synthesis of Z-CRN MAC using a high-temperature process, while Toh *et al.* [30] reported the synthesis of a Z-CRN-with-nanocrystallites MAC using a low-temperature process. Toh *et al.* also used a Monte-Carlo scheme to simulate the growth process and produced a structure similar to their atomic-resolution image obtained by scanning-transmission-electron-microscopy. These authors further contend that the images in Ref. [29] do not have sufficient resolution to discount the presence of crystallites, but do not rule out the possible fabrication of a pure-Z-CRN MAC. The overall conclusion from the above analysis of prior work is that the most likely structure of as-synthesized MAC is a Z-CRN containing nanocrystallites, but one cannot rule out the possibility that a synthesis method can be devised that yields a pure Z-CRN.

On the other hand, monolayer amorphous BN (ma-BN) has not been fabricated so far. A single simulation has been reported, based on disordering crystalline monolayer h-BN by introducing defects, but the overall atomistic structure was not resolved [31]. Three-dimensional amorphous BN (a-BN) in both bulk and thin-film forms, has a long history and multiple uses. It has been found that a-BN consists almost exclusively of $sp^2$ bonding [32,33]. This fact and the fact that both graphene and monolayer h-BN are honeycomb structures consisting exclusively of $sp^2$ bonding, lead us to infer that ma-BN is likely to have a MAC-like structure. Since ma-BN is a binary compound, however, the distribution of the B and N atoms in a MAC-like network remains a big unknown. In particular, if noncanonical B-B and N-N bonds were absent, the structure of ma-BN would not be MAC-like, as five- and seven-member rings must have at least one B-B or N-N bond, instead featuring a mix of only six- and eight-member rings. We note, however, that B-B and N-N bonds are common in compound materials and molecules, e.g., the so-called MAB layered transition-metal borides [34] such as MoAlB, and transition-metal diboranes [35] are known to have strong covalent B-B bonds, while azobenzene [36] features strong N-N bonds. It is clear that there can be more complexity in possible ma-BN structures than in MAC.

In this paper, we adopt the Monte Carlo algorithm introduced for MAC in Ref. [30], carry out simulations of MAC, generalize the algorithm to pursue simulations that mimic the growth of monolayer amorphous BN (ma-BN) from random distributions of B and N atoms in a plane, and compare the evolution characteristics and resulting structures of MAC and ma-BN. The evolution of our systems is governed by processes that lower the energy in ways that mimic the chemical-vapor-deposition (CVD) growth process under different conditions. We find that, unlike the structure of elemental MAC*, the most likely structure of binary ma-BN is altogether different from either of the two well-debated options, pure or crystallite-containing Z-CRN*. In the MAC case, randomly distributed C atoms quickly acquire threefold coordination and form distorted graphene-like crystallites within a network that gradually becomes a CRN, in accord with the simulation results of Ref. [30]. The fabricated MAC corresponds to this stage of the simulation because of the minimal heat that is provided. Continuing the simulation, which corresponds to continuing annealing of the structure, results in growing and merging crystallites, nanocrystalline graphene, and ultimately graphene simply because that is the lowest-energy structure, except possibly for defects that arise from local geometric steric constraints. In sharp contrast, under the same conditions, the simulations yield an altogether different structure for ma-BN. Randomly distributed B and N atoms quickly form a bonded network that comprises canonical B-N bonds and noncanonical B-B and N-N bonds. Threefold coordination ($sp^2$ bonding) is energetically preferred as in h-BN and gradually dominates, but both hexagons and other polygons, with very few exceptions, are not canonical, i.e., they do not feature alternating B and N atoms (they are compositionally disordered). It is the binary nature of BN and concomitant steric constraints plus the viability of noncanonical bonds that reduce significantly the probability of forming canonical hexagons and h-BN-like crystallites, no matter how long the simulation time. On the other hand, because hexagons are lower in energy than all other polygons, we find that “pseudocrystallites”, comprising noncanonical hexagons, form and grow very quickly in ma-BN, much like crystallites form and grow in MAC, but with compositional disorder in the whole structure. All simulations were carried out using empirical potentials that were extensively tested against pertinent density-functional-theory calculations. Most importantly, we find that canonical hexagons and true crystallites do not form in ma-BN unless we artificially increase the energy of canonical bonds by an unrealistic ~22% over the experimentally measured value in h-BN. Our results are corroborated by well-known difficulties to grow h-BN [37-39]. Nevertheless, our prediction that pseudocrystallites are favored in ma-BN the same way as true crystallites are favored in MAC, remains to be tested by experiments similar to those of Ref. [30]. On the other hand, irradiation of monolayer h-BN by an electron beam is likely to produce different forms of disordered BN, while the presence of true nanocrystallites would be controllable by the duration of the irradiation. In fact, this issue may underlie the fact that, unlike the case of graphene, growing h-BN on a substrate requires complex processing. We propose that straightforward CVD processes that produce high-quality graphene on a substrate may produce compositionally disordered Z-CRN ma-BN with pseudocrystallites. Finally, the generality of our conclusions in other amorphous 2D materials and their applicability to 3D materials are discussed.

We employed a kinetic Monte Carlo (kMC) algorithm as in Refs. [30,40,41] to simulate the formation of amorphous materials as an annealing process. First, we performed kMC simulations of the growth of MAC using the potentials of Ref. [42] that were also used in Ref.

[30] and are widely accepted as accurate. As shown in Figure 1, starting with randomly distributed carbon atoms, we first relaxed them using the empirical force field and obtained a disordered C network. We then started the kMC process, whose basic step is a bond rotation as in the simulations that amorphize crystalline graphene [26,27], but *we automatically accept the rotations when they lower the energy* and accept them only when they meet a statistical measure when they raise the energy (see Methods). We found that hexagons and graphene crystallites form quickly, surrounded by a disordered carbon network, as shown in Figure 1c, evidently because hexagons have lower energies than other structural units [43-48]. The nanocrystals continue to grow as the "annealing" continues and the disordered C network gradually turns to a 2D CRN of threefold-coordinated C atoms as shown in Figure 1d, which is consistent with the crystallite model of MAC [30]. Continuing annealing leads to growing crystallites merging to form polycrystalline graphene and finally single-crystal graphene with some defects (Figures 1e and f). The physical interpretation of this sequence is that low temperatures and short times would produce MAC while high temperatures and/or long times would produce either polycrystalline or single-crystal graphene.

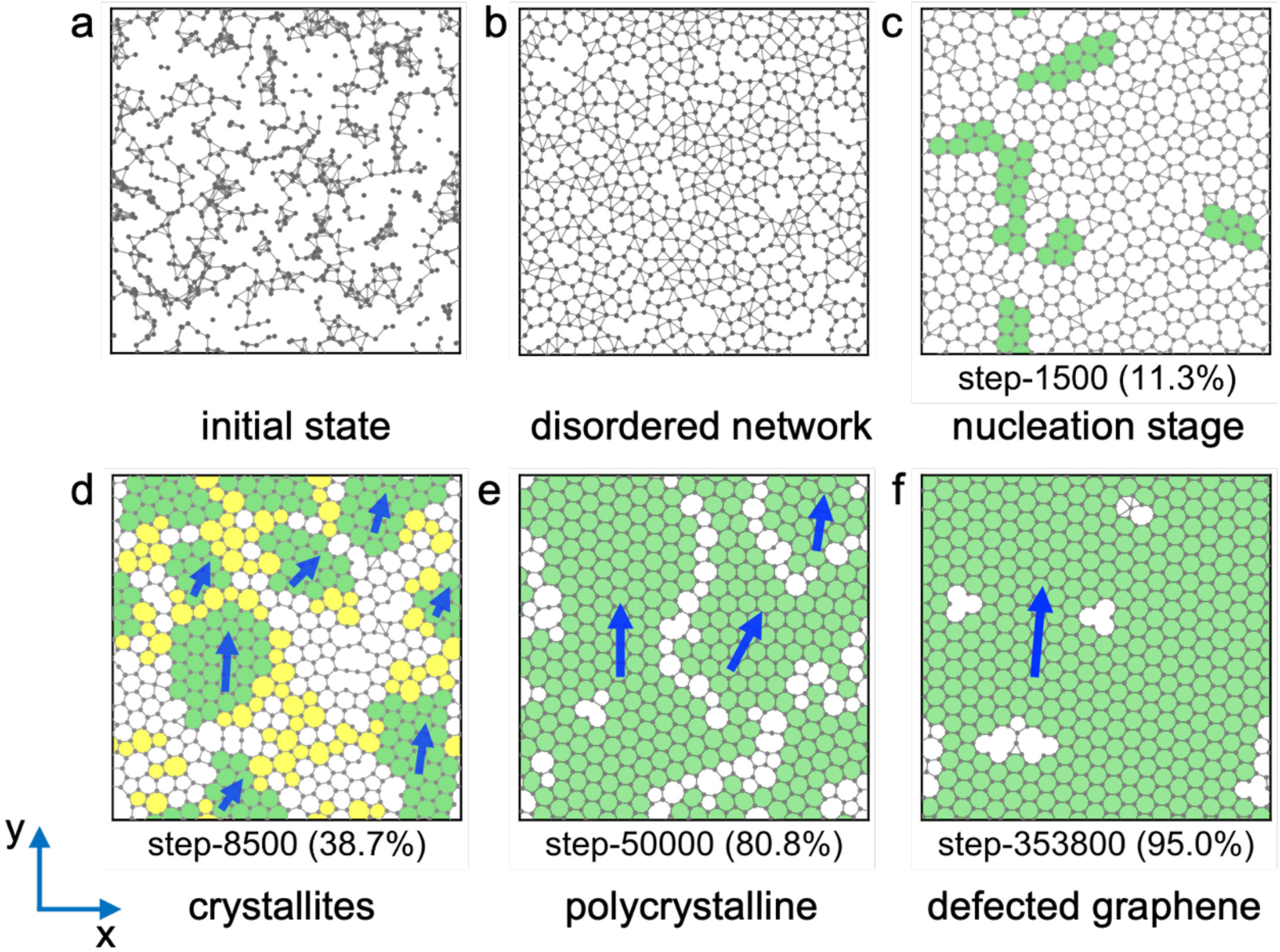

**Figure 1.** Atomic structures of MAC in kMC annealing. (a) The initial configuration with randomly distributed carbon atoms. (b) A disordered network forms after relaxation. (c) Nucleation stage with carbon hexagons, colored green. The percentage of hexagons in crystallites is listed in parentheses. (d) Several nanocrystallites form fairly quickly, with different orientations (blue arrows). (e) Polycrystalline graphene. (f) Single crystalline graphene with several defects. In (d), the polygons that are colored yellow are 5/7 pairs. There exist only small numbers of them in contrast to the models that are generated by amorphizing crystalline graphene via the introduction of random Stone-Wales rotations by bond rotations as in Refs. 25 and 26.

For simulations of the growth of ma-BN, we used the BN-ExTeP empirical potentials, which were tested extensively in Ref. [49]. Additional tests for atomic rearrangements commonly encountered in our algorithm are described in Supporting Information (Tables S1 and S2). We generalized the kMC algorithm to generate atomic structures of ma-BN by additionally allowing exchanges between neighboring atoms, again automatically accepting them only when they lower the energy. Starting from an initial configuration with randomly arranged B and N atoms (Figure 2a), a disordered network forms after relaxation (Figure 2b). Threefold coordination is again favored. As in the case of MAC, the resulting disordered monolayer BN deviates from a nominal CRN, but, in addition, now it contains random B-N, B-B, and N-N bonds. Nevertheless, noncanonical contiguous BN hexagons form early in the annealing process (Figures 2c and d) and serve as nuclei for the formation of what we call h-BN "pseudocrystallites" (Figures 2e-f). Isolated canonical BN hexagons with alternating B-N appear and disappear randomly, while the number of noncanonical hexagons and pseudocrystallites continues to grow. The material surrounding the growing pseudocrystallites, however, appears to be both structurally and compositionally disordered until relatively late in the simulation when it approaches a structure that is only compositionally disordered. In the final analysis, the structure of ma-BN is like that of MAC, but compositionally disordered, i.e, "pseudo-CRN" containing pseudocrystallites.

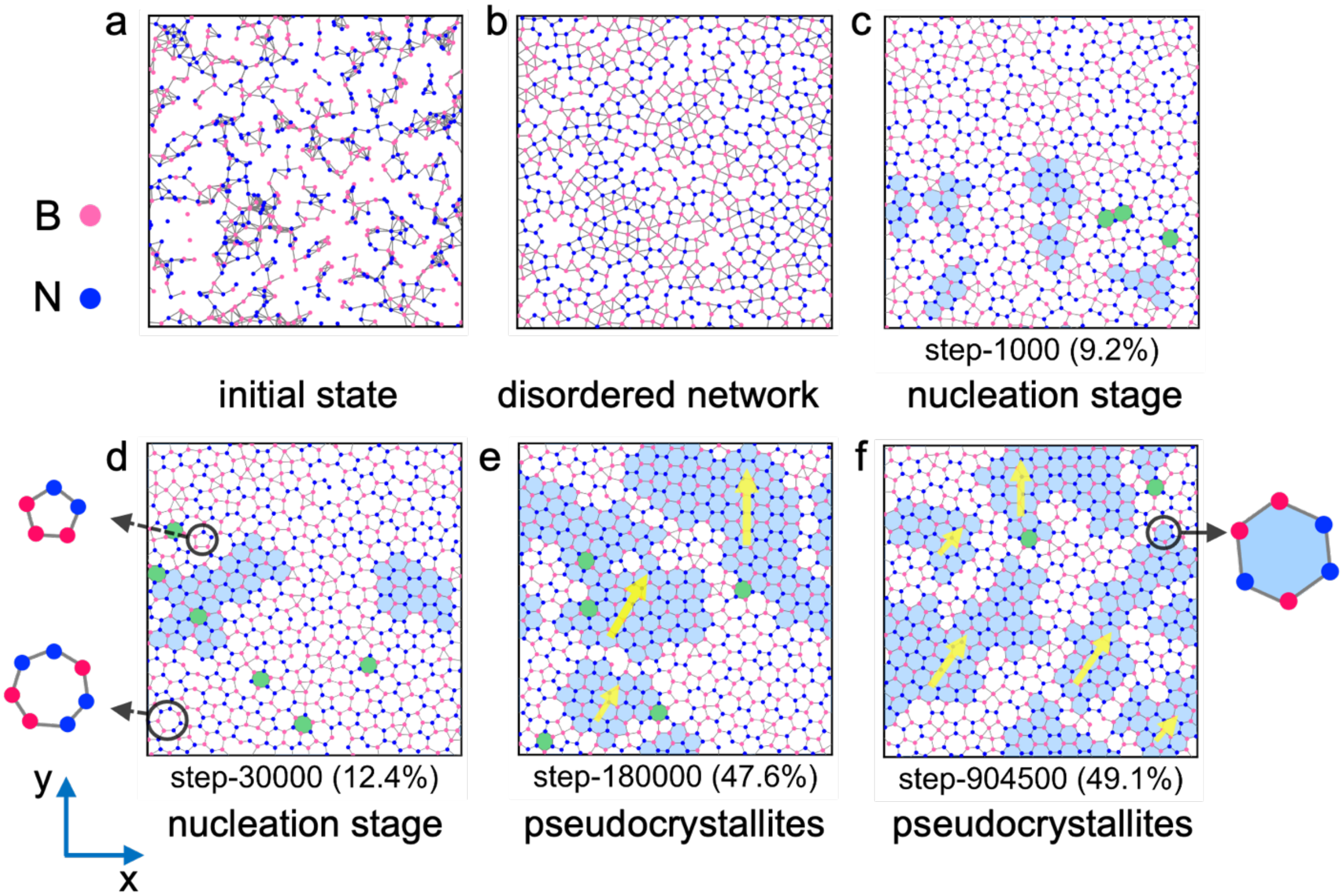

**Figure 2.** Atomic structures of ma-BN in kMC annealing. The h-BN pseudocrystallite regions (compositionally disordered h-BN) are colored blue, and their orientations are marked by yellow arrows. The percentage of noncanonical hexagons in the pseudocrytallites is listed in parentheses. (a) Initial structure with randomly arranged boron and nitrogen atoms. (b) Early stage of disordered BN network. (c) Nucleation stage with emerging hexagons. (d) Still nucleation stage, *i.e.*, the formation of pseudocrystallites in ma-BN is a slower process than the formation of crystallites in MAC. Two typical noncanonical five- and seven-member rings are highlighted in black circles. (e)-(f) Large-area pseudocrystallites comprising noncanonical hexagons (a typical one is highlighted in a black circle), embedded in a pseudo-CRN environment. Note the similarity between the development of the pseudocrystallites in ma-

BN and the crystallites in MAC in Figure 1. After the 180000$^{th}$ step, the structure remains on average the same, with "grain boundaries" containing an alternation of five- and seven-member rings are visible, but the system never evolves into a nano-pseudocrystalline structure. Canonical hexagons (green) remain extremely rare.

The structural difference between MAC and ma-BN originates from the binary nature of BN. The presence of only C atoms in MAC means that there are only C-C bonds, whereby all the hexagons are automatically canonical. The dominant factor in the kMC algorithm for MAC is the energy competition between carbon hexagons and non-hexagons and the former gradually prevails. In the ma-BN case, the hexagon/nonhexagon competition is still at play and gradually hexagons prevail, but there is a background competition between canonical B-N bonding and non-canonical B-B and N-N bonding. *Though B-N bonding is energetically preferred, the formation of B-B and N-N bonds is favored by statistics 10 to 1* (the geometric probability of forming a canonical BN hexagon by three B and three N atoms is $1/10$; see Figure S7). Simulations in which we intentionally enhance the B-N bond energy find that we would need an unrealistic enhancement of more than 1 eV to enable the generation of h-BN crystallites (see Sections II and III of Supporting Information for details of these and other tests of the robustness of our kMC-derived conclusions).

The above conclusions are corroborated by the fact that the growth of crystalline h-BN on substrates is a more complicated and steered process than growing graphene [37-39]. In fact, we propose that a process similar to the growth of MAC may lead to the growth of structurally and compositionally disordered BN (possibly pseudo-CRN as defined above) containing h-BN pseudocrystallites, probably with regions of excess B or N.

The energy evolution of the kMC simulation for ma-BN is shown in Figure 3a. The total energy fully converges after ~180,000 kMC steps. The equilibrium snapshot at the 180,000th step is chosen as the representative configuration of ma-BN in all subsequent discussions. Meanwhile, the thermodynamic stability of ma-BN (step-180000th) is confirmed by long-time 10-ns classical MD simulations at 300 K (Figure 3b). The converged energies fluctuate ~10 meV/atom for both h-BN and ma-BN, smaller than $k_BT$~26 meV at room temperature, reflecting the energetic stability. At the end of the MD simulation, ma-BN ripples to a thickness of 9.5 Å (Figure 3c), while MAC ripples to ~7 Å (align with the experimental values 6-8 Å depending on the substrate [30], see also Figure S8). The large thermal corrugation exhibited in the ma-BN MD simulations does not cause any mechanical instabilities as the monolayer retains its in-plane topography and the pseudocrystallites remain intact (Figure 3c), confirming the structural stability of ma-BN. Pseudocrystallites are also stable when they contain excess B or excess N (see Figure S4).

The evolution of the numbers of three different types of bonds in the kMC is shown in Figure 3d. At the beginning, the three kinds of bonds change and fluctuate, then they reach equilibrium at about the 180,000th step. The bond number ratio of this equilibrium snapshot – num(B-N):num(B-B):num(N-N) ≈ 3:1:1, showing the dominance of B-N bonds. However, a substantial percentage, 39.1%, of the total bonds are noncanonical and are responsible for the formation of the pseudocrystallites and the residual surrounding material that, in the early stages, is both structurally and compositionally disordered. The evolution of structural units is shown in Figure 3e. At the initial stage of the kMC, there are many pentagon-heptagon (5/7) pairs, while the hexagons are rare. With the rapid growth of noncanonical hexagons, the crossover of noncanonical hexagons and 5/7 pairs emerge at the ~20,000th step. Thereafter, the noncanonical hexagons become dominant and saturate at the 180,000th step, where they contribute about 63.7% of all structural units. The subdominant structural units are the 5/7 pairs, whose contribution decreases from ~40% to ~30%. The 5/7 pairs also form boundaries (yellow-orange regions in Figure 4a) that separate the pseudocrystallite regions. The four- and eight-membered rings (4/8) contribute little (~1.6%). The $B_3$ triangles in ma-BN are structural elements of borophene and contribute only 2.7%. The canonical hexagons remain negligible (~1.3%). Thus, the material surrounding the pseudocrystallites becomes essentially a pseudo-CRN as defined earlier. This equilibrium structure, which remains stable after the 180,000$^{th}$ step, is likely to be the structure that is obtained by straightforward CVD without special procedures that are needed to produce crystalline h-BN.

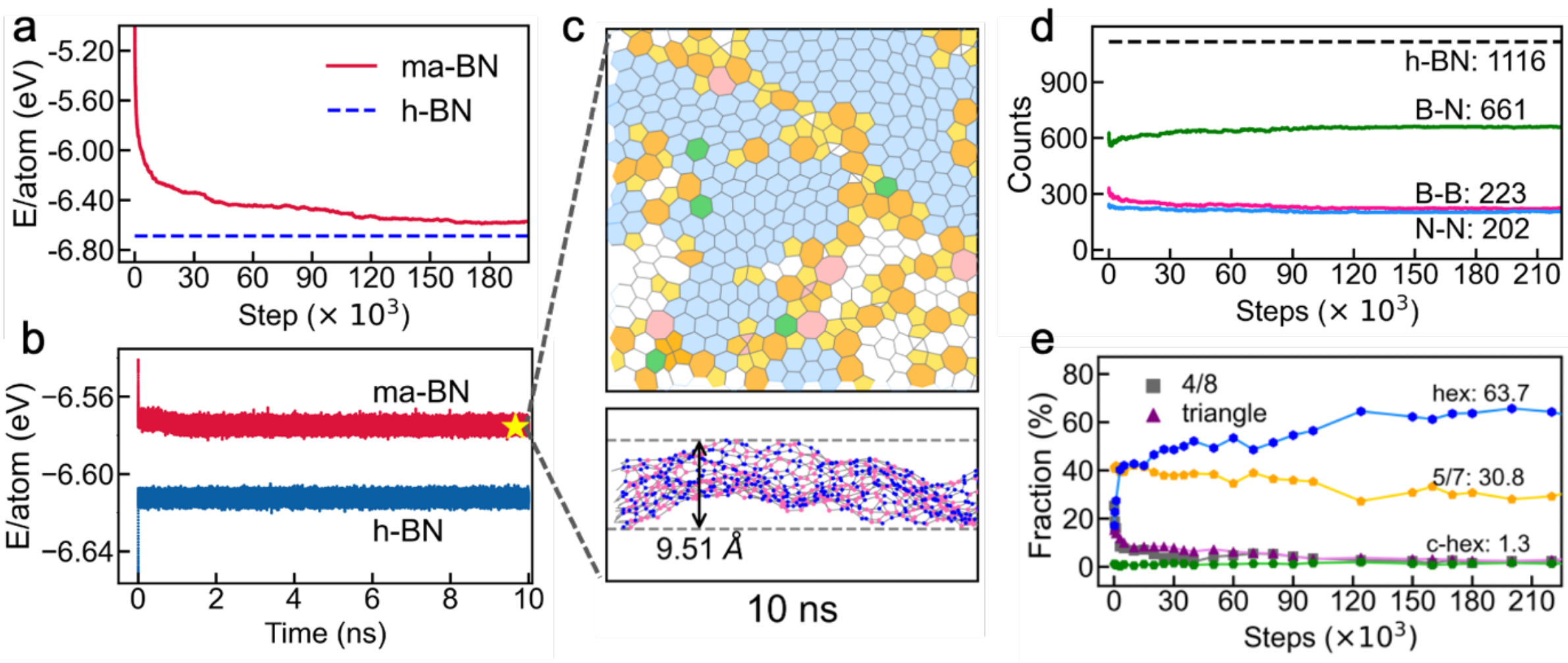

**Figure 3.** The thermal stability of ma-BN and the kMC statistics. (a) Energy evolution in the kMC simulation. (b) Energy evolutions in the classical MD simulations. The equilibrium-stage ma-BN snapshot (step-180000th) was placed in an NPT (isothermal-isobaric) ensemble at 300 K and evolved for 10 ns. (c) Top and front view of the final snapshot of ma-BN after a 10-ns classical MD simulation at 300 K. Note the initial configuration (step-180000th ma-BN, see Figure 2e) is flat. (d) Evolution of counts of different bond species during the kMC simulation. The bond numbers at the 180000th snapshot are listed. The reference black dashed line represents the number of B-N bonds in a same-sized h-BN supercell (equals 1116). (e) Counts of structural units during the kMC simulation. Canonical hexagons (c-hex) are colored in green, while noncanonical hexagons are in blue.

We also studied the electronic properties of ma-BN. Figure 4b shows the electronic density of states (DOS) of the equilibrium stage ma-BN (step-180000th). Different from h-BN, which has a large band-gap, ma-BN features a continuum of in-gap states. However, the charge densities of the gap states near the Fermi energy are localized and disconnected, mainly at the boundaries formed by 5/7 pairs (see Figure 4a), signaling the presence of a transport gap, i.e., ma-BN is an insulator. Since the states near the Fermi energy are very similar to those in MAC, however, its resistivity may show a similar power-law behavior, namely $\rho \sim T^N$. *In contrast, h-BN is a wide-gap semiconductor that can be doped n-type or p-type.*

The calculated radial distribution function (RDF) (Figure 4c) of the equilibrium stage ma-BN (step-180000th) shows that the middle- and long-range RDF peaks are highly broadened or missing, while the short-range peaks are basically preserved, establishing ma-BN to be a truly 2D amorphous material. Specifically, the first RDF peak at around 1.5 Å represents the nearest-neighbor coordination. N-N and B-N bonds contribute to the first subpeak at 1.41 Å, while B-B bonds contribute to the second subpeak at 1.54 Å (Figure 4d). The broad subpeaks reveal bond-length variations. The dominant threefold coordination (Figure 4e) is driven by $sp^2$ bonding. A broad peak near 120° in the bond angle counting (Figure 4f) reflects the existence of a substantial number of distorted noncanonical hexagons and other polygons. The threefold coordination and bond length/angle variations demonstrate that the ma-BN features distorted $sp^2$-bonding.

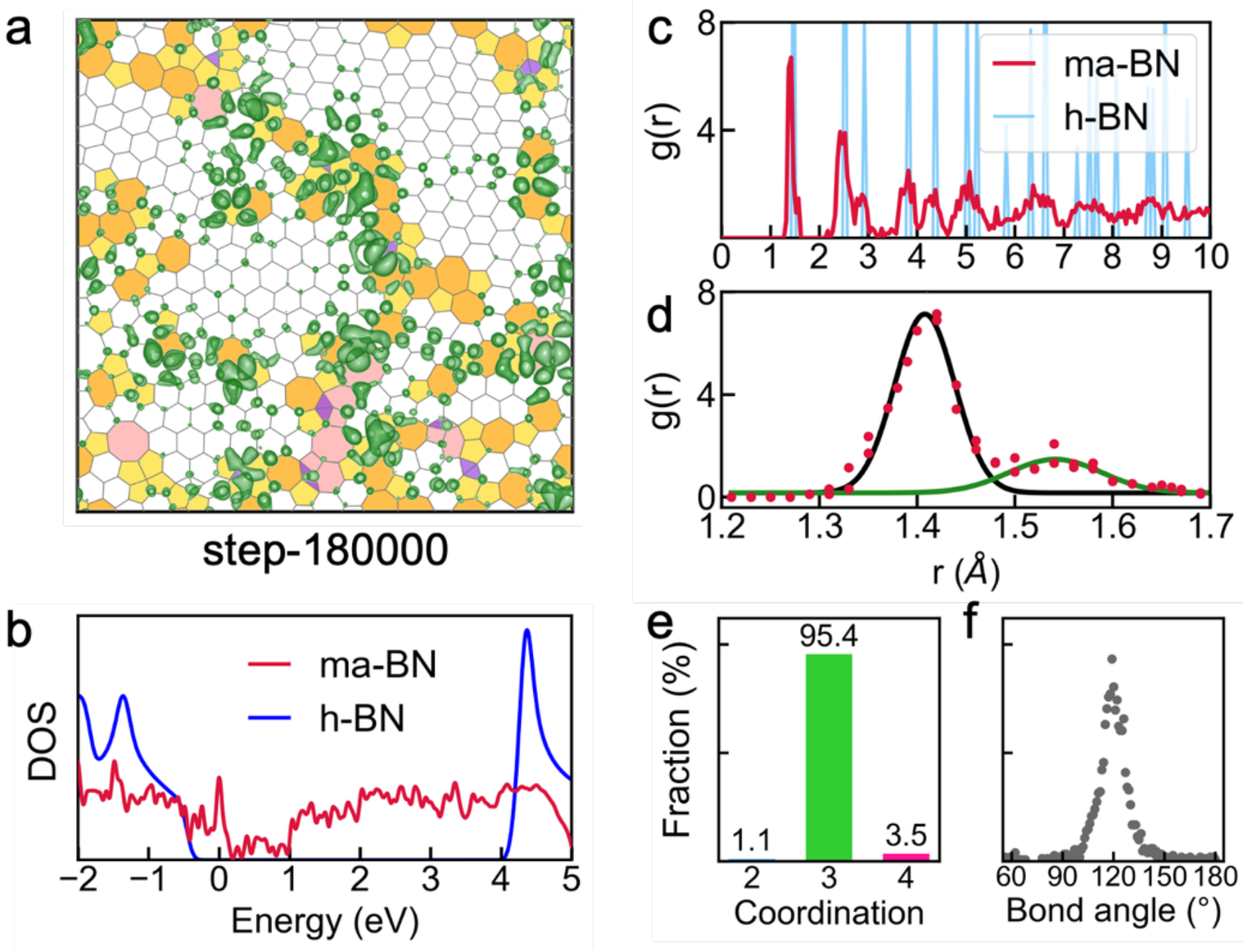

**Figure 4.** Electronic and structural properties of ma-BN. The equilibrium snapshot is chosen as in Figure 2e. (a) Charge density distribution in real space in the energy range $E_F \pm 0.2$ eV. (b) Calculated DOS of ma-BN and h-BN, the Fermi energy is at 0 eV. (c) RDF of ma-BN compared with h-BN. (d) Zoom-in plot of the first RDF peak. (e) Counts of coordination numbers. (f) Counts of bond angles.

In summary, we predicted that the atomic structure of amorphous monolayer BN is something altogether different from the two canonical options, namely Z-CRN or nanocrystallite-containing Z-CRN. We find that, depending on the growth method, the most likely structure is a compositionally disordered, pseudo-CRN that is likely to contain compositionally disordered pseudocrystallites, which are a unique feature in binary and multi-element amorphous materials. The difference with monolayer amorphous carbon is caused by the low probability of forming canonical BN hexagons imposed by the additional steric constraint created by the two atomic species. The ma-BN is structurally and thermodynamically stable at room temperature. As an insulating monolayer with good stability at room temperature, ma-BN may be suitable for similar applications as monolayer amorphous carbon [29,30]. Finally, since it is already known that bulk a-BN is almost exclusively $sp^2$-bonded, as we predict to be the case for ma-BN [32,33], it appears that bulk a-BN may be tangled ma-BN sheets, but we cannot tell for sure. Since the B/N atomic ratio was found to be about 1:1.08 [33], it is certain that there are wrong bonds and possibly pseudocrystallites.

The generality of the ma-BN results to other binary-compound monolayers and the implications for other two- and for three-dimensional amorphous compounds cannot be fully resolved at this point. The formation of noncanonical bonding in a binary monolayer is favored by statistics but may not be favored by bond energetics. However, due to the lack of adequately reliable empirical potentials for some of the most typical monolayer materials like borophene, monolayer GaAs, and monolayer BeO, our conclusions on the distinction between elemental and binary amorphous materials remains to be confirmed. Our simulation results, however, suggest that compound amorphous 2D or 3D materials have more options than the binary choice of Z-CRN or crystallite-containing Z-CRN. Network glasses like a-$SiO_2$, are unlikely to exist in a form that features noncanonical O-O and Si-Si bonds. Other materials such as bulk amorphous BN, SiC, and amorphous III-V semiconductors, however, may form either Z-CRN or pseudo-CRN structures with or without crystallites or pseudocrystallites, depending on the method of preparation. As we noted in the introduction, amorphous BN is known to have almost exclusively $sp^2$ bonding [32,33], whereby ma-BN-like structures may well be the dominant feature.

ASSOCIATED CONTENT

**Supporting Information**

The Supporting Information is available free of charge.

Details of kMC simulations, molecular dynamics, DFT calculations; accuracy tests of the classical potential for ma-BN; extended numerical tests of kMC results including adding constraint, varying kMC acceptance rate, tuning chemical stoichiometries, estimating canonical bond forming with DFT, and simulating with reax force field; additional data of bonding probability, thermal ripples of MAC and ma-BN. (PDF)

**Notes**

The authors declare no competing financial interest.

ACKNOWLEDGMENT

We acknowledge financial support from National Key R&D program of China (Nos. 2019YFA0308500) and the K. C. Wong Education Foundation. Work at Vanderbilt was funded by the U.S. Department of Energy, Office of Science, Basic Energy Sciences, Materials Science and Engineering Division grant no. DE-FG02-09ER46554 and by the McMinn Endowment.